\documentclass[printer]{raa}            % referee version: for submission

\usepackage{graphicx,times}             %for PS/EPS graphics inclusion, new

\begin{document}

   \title{A Method for Calculating Collision Probability Between Space Objects
%\footnotetext{$*$ Supported by the National Natural Science Foundation of China.}
}
%   \subtitle{I. Place Your Subtitle Here}

   \volnopage{Vol.0 (200x) No.0, 000--000}      %%preserved for Editor. DOn't remove!
   \setcounter{page}{1}          %%starting page, preserved for Editor. DOn't remove!

   \author{Xiao-Li Xu
      \inst{1,2}
   \and Yong-Qing Xiong
      \inst{1,2}
   }

   \institute{Purple Mountain Observatory, Chinese Academy of Sciences, Nanjing 210008, China; {\it xlxu@pmo.ac.cn}\\
        \and
             Key Laboratory of Space Object and Debris Observation, Chinese Academy of Sciences, Nanjing 210008, China\\
   }

   \date{Received~~2013 month day; accepted~~2013~~month day}

\abstract{ A method is developed to calculate collision probability in this paper.
Based on the encounter geometric features of space objects, it is reasonable to
separate the radial orbital motions from that in the cross section for most encounter events in near circular orbit.
Therefore, the collision probability caused by orbit altitude difference in the radial direction and the collision
probability caused by arrival time difference in the cross section are calculated respectively.
The net collision probability is expressed as an explicit expression by multiplying the above two components.
Numerical cases are applied to test this method by comparing the results with the general method. The results indicate that this method is valid
for most near circular orbital encounter events.
}

   \authorrunning{X.L. Xu,\& Y.Q. Xiong }            %author_head in even pages
   \titlerunning{A Method for Calculating Collision Probability Between Space Objects }  % title_head in odd pages

   \maketitle
%% The author head (on even pages) and the title head (on odd pages) will be
%% automatically extracted from \author{} and \title{}. Whenever the title is too long,
%% you will be asked to supply a shorter one by inserting either \authorrunning{} or
%% \titlerunning{} before \maketitle. Anyway, you can specify your own heads.
%%
%%
%% Note: In the following text body of your manuscript, please note several differences from
%%       other major journals:
%% (1) \subsection{Please Capitalize the First Letter of Each Notional Word in Subsection Title}
%% (2) Please Capitalize the First Letter of Each Notional Word in all tables' captions

%
%________________________________________________ sections below
%
\section{Introduction}           %% first-level sections will be auto-capitalized
\label{sect:intro}

Now a very large number of space debris are orbiting in near earth orbit below 2000\,km;
they could potentially threaten space missions and spacecrafts(Sun \& Zhao~\cite{sz2013}).
Collision events could disturb/damage spacecrafts as well as produce new debris, leading to a vicious cycle of space environment.
Although a collision event between two space objects barely happens,
several collision accidents have already been confirmed so far.
In order to prevent space collision events,
effective collision risk assessment and avoidance maneuver have been carried on (Akella \& Alfriend~\cite{kl2005};  Francois \& Eloy~\cite{fe2008}).

The earliest risk criterion was the exclusion volume method (Leleux et al.~\cite{ls2002}). An exclusion volume surrounding
the spacecraft is specified, and an alarm is raised if any object penetrates the exclusion volume.
However, this method is too conservative to raise a high number of false alarms.
Therefore, the probability threshold method was proposed (Kelly \& Picciotto~\cite{kp2005}); when a probability threshold is set,
an alarm is raised as the calculated collision probability exceeds this threshold.
This approach reduces false alarms and thus is more rational than the exclusion volume method.
Many efforts have been dedicated to the effective formulations of collision probability calculation,
and significant progress have been made in recent years. To calculate collision probability between two objects,
covariance models for both should be applied.
It is permissible to combine the error covariance matrices for two orbiting objects to obtain a relative
covariance matrix as long as they are represented in the same frame (Chan~\cite{ch1997}).
The combined covariance matrix has an associated three-dimensional probability density function (PDF)
that represents the relative position uncertainty of the two objects.
Since most encounters between space objects are occurring at high relative velocities,
the collision probability calculation can be reduced to a two-dimensional integral over a circular
region in a plane normal to the relative velocity vector referred to as the encounter frame (Akella \& Alfriend~\cite{aa2000}).
Patera (~\cite{pa2001}) further developed an approach, which reduces the calculation to a single closed-path integral,
to compute collision probability between two irregular objects.
Bai \& Chen (~\cite{bc2008, bc2009}) also developed one method  based on space compression and coordinate rotation,
in which the collision probability is expressed as an explicit function of the encounter geometry.
At present, the probability threshold method has been widely adopted to assess collision risks for space objects.

The general method for calculating collision probability involves a projection from a three-dimensional PDF to a two-dimensional one.
The initial state vectors of two objects in the inertial frame are given by $\mathbf{r}_1,\mathbf{r}_2$,
and their error covariance matrices in the local frame are defined as $\mathbf{C}_1,\mathbf{C}_2$.
The transformation matrices from the local to the inertial frame are given by  $\mathbf{P}_1$ and $\mathbf{P}_2$, respectively.
The respective error covariance matrices in the local frame are transformed to the ones in the inertial frame in the usual manner:
\begin{equation}
\begin{array}{ll}
\mathbf{C}_{1I} =& \mathbf{P}_1\mathbf{C}_1\mathbf{P}_1^{-1} \,,\\
\mathbf{C}_{2I} =& \mathbf{P}_2\mathbf{C}_2\mathbf{P}_2^{-1} \,.
\end{array}
\end{equation}

Because the position errors of two objects are uncorrelated,
the error covariance matrices are combined to obtain the relative error covariance matrix in the inertial frame:
\begin{equation}
\mathbf{C}_I=\mathbf{C}_{1I}+\mathbf{C}_{2I} .
\end{equation}

Then the relative position and velocity are needed to define the encounter frame
whose $z$-axis is parallel to the relative velocity vector.
The transformation from the inertial to the encounter frame given by $\mathbf{U}$ is used to
derive the relative position vector and error covariance matrix in the encounter frame:
\begin{eqnarray}
\Delta\mathbf{r}_e &=& \mathbf{U}\mathbf{r}_2-\mathbf{U}\mathbf{r}_1 \,,\nonumber\\
    \mathbf{C}_{e} &=& \mathbf{U}\mathbf{C}_I\mathbf{U}^{-1} \,.
\end{eqnarray}
The collision probability in the encounter frame is expressed as:
\begin{equation}
P=\frac{1}{\sqrt{{(2\pi)^3|\mathbf{C}_e|}}}\int\int\int_V\rm{exp}\{
-\frac{1}{2}
\Delta\mathbf{r}_e^{\rm{T}}\mathbf{C}_e^{-1}\Delta\mathbf{r}_e\}\rm{d}V \,.
\end{equation}

Because most encounters have high relative velocities,
the relative position vector is orthogonal to the relative velocity vector when two objects are at the closest point of approach (CPA)(Chan~\cite{ch1997}).
Therefore, the collision probability calculation can be reduced to a two-dimensional integral over a circular region in the encounter frame:
\begin{equation}
P=\frac{1}{2\pi\sqrt{|\mathbf{C}_e^{\prime}|}}\int\int_{x^2+y^2\leq{R^2}}\rm{exp}\{
-\frac{1}{2}
\Delta\mathbf{r}_e^{\prime\rm{T}}\mathbf{C}_e^{\prime-1}\Delta\mathbf{r}_e^{\prime}\}\rm{d}x\rm{d}y \,.
\end{equation}

From Eq.~(5), we can see that collision probability depends on the relative position,
error covariance matrix and object's size. It has to be calculated numerically in light of this implicit expression, and some geometric features of encounter are implied.
The general method for collision probability provides sufficiently accurate results, but does not reveal the direct connection between collision probability and conjunction geometry.

Since the sensitivity of collision probability to conjunction is significant for making avoidance maneuver strategies, an explicit expression of collision probability in terms of relevant
parameters is needed. The purpose of this work is to provide a method to calculate collision probability expressed as an explicit function.
Based on the formulate of the general method and the geometric features of encounter described in section~2.1  ,
this approach will separately consider the collision probability caused by orbital altitude difference in the radial direction (section~2.2.1),
and the collision probability caused by arrival time difference in the cross section of the two objects (section~2.2.2).
Then the net collision probability is obtained by multiplying the above two probabilities (section~2.2.3).
Numerical cases are listed to test this method in section~3, and the conclusions are given in section~4.

%% Authors can give a citation as 'Michel et al. 1992'.
%% You may also use \cite, \citep and \citet for citation, and use Table~1 or Figure~1
%% and so forth. Using \ref and \label for cross-references of Tables/Figures
%% is a good way in adjusting/adding/removing text, tables or figures.

\section{Methodology}
\label{sect:Meth}

\subsection{Background and foundation}

According to the geometric features of encounter as shown in Fig.~1,
a collision between two space objects could only happen near the intersection of their orbit planes.
Therefore, the orbital altitude difference and arrival time difference when they are passing by the
intersection of their orbit planes can be used to filter dangerous encounters.
The orbital altitude difference is along the radial direction, and the arrival time difference is determined by the velocities of motion in the cross section.
If the cross section is perpendicular to the radial direction, thus the relative motion of two space objects can be divided to two independent parts,~i.e., the radial motion and the motion within the cross section. Such a prerequisite is naturally satisfied for the encounter of two space objects in near circular orbit.
\begin{figure}[htbp]
  \centering
   \includegraphics[width=0.6\textwidth]{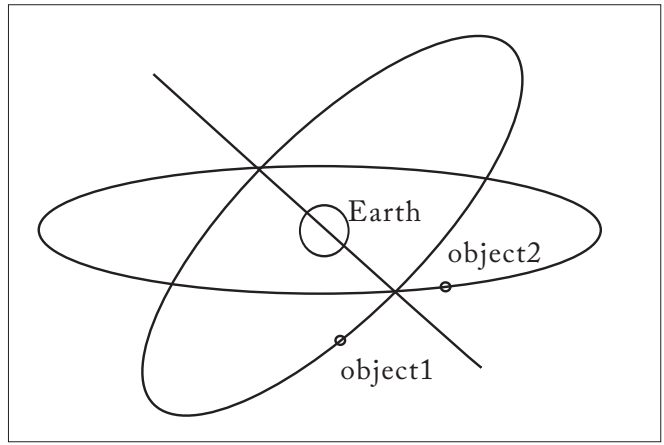}
  \caption{ \footnotesize Encounter geometry of space objects. When a close encounter happens, the two objects are both orbiting near the intersection of their orbit planes.}
  \end{figure}

Hence, the eccentricity distribution of space objects shows the necessity to be discussed. For our space missions and spacecrafts, dangerous objects in low-earth-orbit are to our most concern. Therefore, we select 13,000 space objects with altitudes of the perigee below 2000\,km from NORAD Two-Line Element (TLE) Sets. The eccentricity distribution for these space objects is shown in Fig.~2.
The overall eccentricity distribution is dominated by the small eccentricities, i.e., more than $90\%$ space objects are in near circular orbits with eccentricities less than 0.1.
Moreover, the eccentricity distribution also varies with altitude of the perigee, the fraction of large eccentricities increases with decreasing altitude.
There are a few numbers concentrated in large eccentricity distribution ($0.5\sim0.8$), whose altitudes of the orbit perigee are only several hundred kilometers ( mostly between 200km and 800km) as shown in the left subplot. For these objects in high eccentricity orbit ( HEO ), they move faster and pass less time near the perigee according to the motion feature of elliptical orbit.
Therefore, most encounters are happening between two space objects in near circular orbit.
\begin{figure}[htbp]
  \centering
   \includegraphics[width=1.0\textwidth]{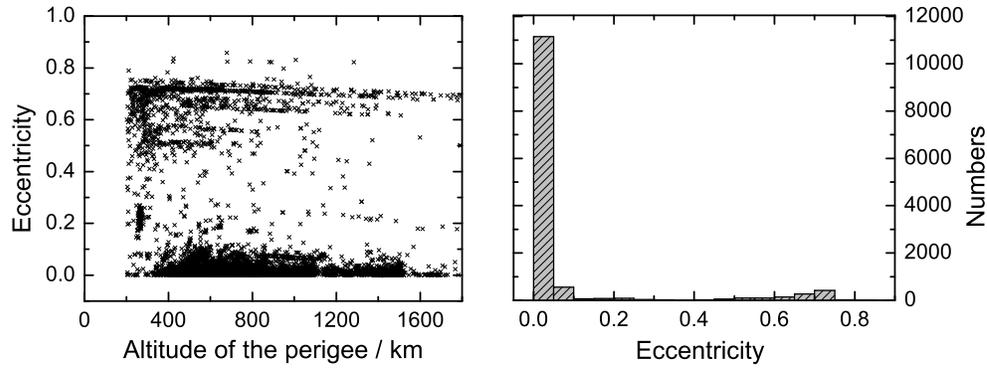}
  \caption{ \footnotesize Eccentricity distribution of space objects in near earth orbit below 2000\,km. \textit{Left}: Eccentricity distribution
  versus the altitude of the orbit perigee.\textit{ Right}: Statistics for eccentricity distribution. }
  \end{figure}

 \begin{figure}[htbp]
  \centering
   \includegraphics[width=1.0\textwidth]{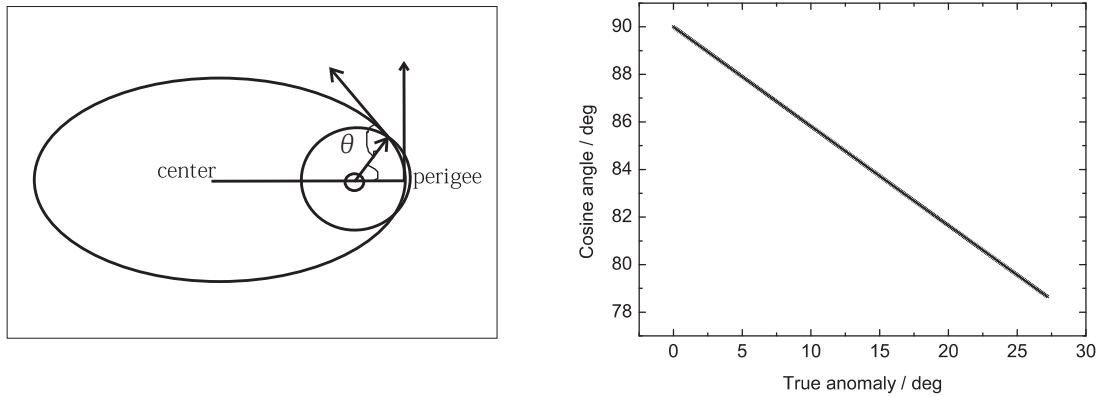}
  \caption{ \footnotesize Orthogonality between the radial direction and the motion direction for dangerous objects in larger eccentricity orbit.
  \textit{Left}: Close encounter geometry for a satellite in a circle orbit with a dangerous object in a large eccentricity.
  The eccentricity is set as 0.7. The cosine angle $\theta$ is defined as the inclined angle between the radial direction and the cross section.
  \textit{Right}: The cosine angle between the radial direction and the velocity direction varies with the true anomaly. }
  \end{figure}

However, a close encounter between a near circular orbit below 2000\,km and
a high eccentricity orbit could only happen near the perigee.  A sketch of such case is shown in the left subplot of Fig.~3; the eccentricity of the elliptical orbit is 0.7.
Obviously, at the point of the perigee ($f=0^\circ$), cosine angle $\theta=90^\circ$, the radial direction is orthogonal to the cross section.
The right subplot of Fig.~3 shows that cosine angle $\theta$ varies with true anomaly $f$;
$\theta$ equaling to $90^\circ$ indicates vertical, and it can reach a extreme value about $12^\circ$ deviation from vertical in this case.
Actually, dangerous encounters between space objects in high eccentricity orbit just like this case rarely occur. Moreover, we also assess encounters between these objects.
A geometric filter with altitude difference less than 10\,km and time difference within 15\,s is set to screen dangerous encounters.
Among these dangerous encounter events, the direction cosine angles between the radial direction and the cross section are also calculated, and they are consistently approximating to $90^\circ$.

Through the above analysis, it is reasonable to separate orbit motions in the radial direction from that in the cross section for most close encounter events in near circular orbit.
Based on this premise, we can separate the net collision probability into two parts: one is in the radial direction which is caused by the orbital altitude difference,
the other is in the cross section which is caused by the arrival time difference. The net collision probability is derived by multiplying the two of them.
The method for calculating collision probability is appropriate for most near circular orbit encounter events.

\subsection{Calculation Algorithms}

\subsubsection{The collision probability in the radial direction}

   \begin{figure}[htbp]
  \centering
   \includegraphics[width=0.5\textwidth]{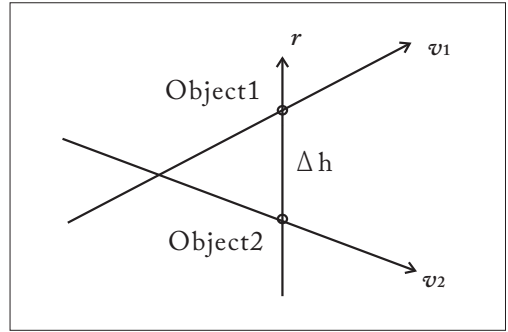}
  \caption{ \footnotesize The relative motion in the radial direction. When the two objects are arriving at the intersection of their orbit planes, the orbital altitude difference
  is set to $\Delta h$. }
  \end{figure}

Firstly, if assuming two space objects arriving at the intersection of their orbit planes at the same time,
one has to only consider the relative trajectory and position error in the radial direction as shown in Fig.~4.
The uncertainty in the relative radial position is defined by an one-dimensional Normal distribution of the form :
\begin{equation}
\rho(r)=\frac{1}{2\sqrt\pi\sigma_r}\rm{exp}\{
\frac{-(r-\Delta h)^2}{2\sigma_r^2}\} \,,
\end{equation}
where $\Delta h$ is the altitude difference of the two objects, and the relative position error in the radial direction is given by $\sigma_r^2=\sqrt{\sigma_{1r}^2+\sigma_{2r}^2}$.

Then the collision probability density distribution function can be given by:
\begin{equation}
F(r)=\frac{1}{2\sqrt\pi\sigma_r}\int_{-\propto}^{r}\rm{exp}\{
\frac{-(r-\Delta h)^2}{2\sigma_r^2}\}\rm{d}r \,.
\end{equation}

It can be substituted by an approximate expression just like:
\begin{equation}
F(r)=\frac{\rm{exp}[a(r-\Delta h)]}{1+\rm{exp}[a(r-\Delta h)]} \,, where \,,
a=\frac{4}{\sqrt{2\pi}\sigma_r} \,.
\end{equation}
The result from this approximate expression is accurate enough without introducing error greater than $2\%$ (Liu et al.~\cite{lu2009}).

Applying the combined size of the objects in the radial direction as $r_a$  , the collision probability caused by altitude difference can finally be expressed as:
\begin{eqnarray}
P_R & = & F(x)\mid_{-\propto}^{r_a} - F(x)\mid_{-\propto}^{-r_a}\nonumber\\
    & = & \frac{\rm{exp}[\frac{4(r_a-\Delta h)}{\sqrt{2\pi}\sigma_r}]}{1+\rm{exp}[\frac{4(r_a-\Delta h)}{\sqrt{2\pi}\sigma_r}]}-\frac{\rm{exp}[\frac{4(-r_a-\Delta h)}{\sqrt{2\pi}\sigma_r}]}{1+\rm{exp}[\frac{4(-r_a-\Delta h)}{\sqrt{2\pi}\sigma_r}]} \,.
\end{eqnarray}

The collision probability in the radial direction is reduced to an analytical expression of radial altitude difference, relative error and combined size.

\subsubsection{The collision probability in the cross section}

The state vectors when the two objects reach the intersection of their orbit planes are labeled as: $\mathbf{r}_1,\mathbf{r}_2$, and $\mathbf{v}_1,\mathbf{v}_2$.
We use a scale factor $\beta=r_1/r_2$ to eliminate the radial altitude difference of the two objects, and thus only consider their relative motion in the cross section.
    \begin{figure}[htbp]
  \centering
   \includegraphics[width=0.8\textwidth]{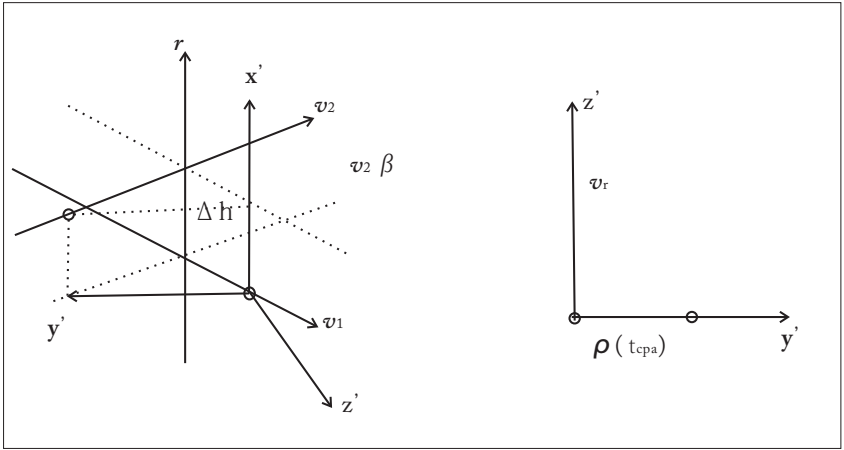}
  \caption{ \footnotesize The reconstructed encounter frame. The $x^{\prime}$ direction is normal to the cross section.
The $y^{\prime}$ direction is defined as the closest approach position vector in the cross section.
The $z^{\prime}$ vector completes the right-handed system.  }
  \end{figure}

In order to compute collision probability in the cross section, the encounter coordinate frame is reconstructed as shown in Fig.~5.
The $x^{\prime}$ direction is normal to the cross section which is defined by the plane containing vectors of $\mathbf{v}_1$ and $\mathbf{v}_2\beta$.
The $y^{\prime}$ direction is defined as the closest approach position vector in the cross section.
The $z^{\prime}$ vector completes the right-handed system. The $o^{\prime}-x^{\prime}y^{\prime}z^{\prime}$ system is centered on one object.

In the cross section,  if the initial position and velocity vectors of two objects are respectively $\mathbf{r}_{10},\mathbf{r}_{10}$,
and $\mathbf{r}_{20}\beta,\mathbf{r}_{20}\beta$. The relative position vector between them is given by:
\begin{equation}
\vec{\rho}(t)=(\mathbf{r}_{20}\beta-\mathbf{r}_{10})+(\mathbf{r}_{20}\beta-\mathbf{r}_{10})t=\vec{\rho}_0+\mathbf{r}_rt \,.
\end{equation}

When the two objects are arriving at the closest point of approach, the equation $\frac{\rm{d}}{\rm{d}t}[\vec{\rho}(t)\cdot\vec{\rho}(t)]=0$ must be satisfied.
We obtain the time of closest approach (TCA):
\begin{equation}
t_{cpa}=-\frac{\vec{\rho}_0\cdot\mathbf{v}_r}{\mathbf{v}_r\cdot\mathbf{v}_r} \,.
\end{equation}

Then the closest position vector is given by:
\begin{equation}
\vec{\rho}(t_{cpa})=\vec{\rho}_0+(-\frac{\vec{\rho}_0\cdot\mathbf{v}_r}{\mathbf{v}_r\cdot\mathbf{v}_r})\cdot\mathbf{v}_r \,.
\end{equation}

Therefore, unit vectors of the reconstructed encounter frame can be defined by:
\begin{equation}
\begin{array}{ll}
\hat{x}^\prime =& \frac{\mathbf{v}_1\times\mathbf{v}_2\beta}{|\mathbf{v}_1\times\mathbf{v}_2\beta|} \,,\\
\hat{y}^\prime =& \frac{\vec{\rho}(t_{cpa)}}{|\vec{\rho}(t_{cpa)}|} \,,\\
\hat{z}^\prime =& \hat{x}^\prime\times\hat{y}^\prime \,.\\
\end{array}
\end{equation}

Based on the foundation described in section 2.1, the $\hat{x}^\prime$ axis is basically parallel to the radial direction in the intersection of orbit planes. The transformation from the inertial frame to the reconstructed encounter frame is built to separate the radial motion from the cross section by directly eliminating motion in the $x^{\prime}$ direction .
At the same time, there is no uncertainty in the direction of relative velocity at TCA in the cross section, so the relative position error can be projected into the direction of the closest approach position vector. After projecting the combined size of the objects into the closest approach position vector direction, the collision probability caused by arrival time difference in the cross section is given by:
\begin{equation}
P_T=\frac{1}{2\sqrt\pi\sigma_t}\int_{-r_t}^{r_t}\rm{exp}\{
\frac{-(y-\rho_{min})^2}{2\sigma_t^2}\}\rm{d}y \,,
\end{equation}
where $\sigma_t$ is defined as the relative position error along the direction of the closest approach position vector,
and it only depends on position errors in the directions of along-track and out-normal.
$\rho_{min}$ is the closest approach distance in the cross section. $r_t$ is the combined size along the direction of the closest approach position vector in the cross section.

Similarly by using the approximate expression we can also obtain an analytical solution:

\begin{eqnarray}
P_T & = & F(x)\mid_{-\propto}^{r_t} - F(x)\mid_{-\propto}^{-r_t}\nonumber\\
    & = & \frac{\rm{exp}[\frac{4(r_t-\rho_{min})}{\sqrt{2\pi}\sigma_t}]}{1+\rm{exp}[\frac{4(r_t-\rho_{min})}{\sqrt{2\pi}\sigma_t}]}
    -\frac{\rm{exp}[\frac{4(-r_t-\rho_{min})}{\sqrt{2\pi}\sigma_t}]}{1+\rm{exp}[\frac{4(-r_t-\rho_{min})}{\sqrt{2\pi}\sigma_t}]} \,.
\end{eqnarray}

\subsubsection{The net collision probability as an explicit function}

In the above two subsections, we separately obtain analytical expressions of collision probability caused by altitude difference and time difference. Collision events may happen only when the two probabilities are big enough at the same time.
Therefore, the net collision probability between two space objects can finally be defined by:
\begin{eqnarray}
P & = & P_R\cdot P_T\nonumber\\
  & = & \{\frac{\rm{exp}[\frac{4(r_a-\Delta h)}{\sqrt{2\pi}\sigma_r}]}{1+\rm{exp}[\frac{4(r_a-\Delta h)}{\sqrt{2\pi}\sigma_r}]}
        -\frac{\rm{exp}[\frac{4(-r_a-\Delta h)}{\sqrt{2\pi}\sigma_r}]}{1+\rm{exp}[\frac{4(-r_a-\Delta h)}{\sqrt{2\pi}\sigma_r}]}\}\nonumber\\
  &\cdot&\{\frac{\rm{exp}[\frac{4(r_t-\rho_{min})}{\sqrt{2\pi}\sigma_t}]}{1+\rm{exp}[\frac{4(r_t-\rho_{min})}{\sqrt{2\pi}\sigma_t}]}
        -\frac{\rm{exp}[\frac{4(-r_t-\rho_{min})}{\sqrt{2\pi}\sigma_t}]}{1+\rm{exp}[\frac{4(-r_t-\rho_{min})}{\sqrt{2\pi}\sigma_t}]}\} \,.
\end{eqnarray}

This method gives an explicit expression to compute collision probability. It is advantageous to analyze the influencing factors of the
collision probability directly. Additionally, the method separates orbit motions in the radial direction from that in the cross section.
Therefore, position errors and geometric sizes of the two components can be differently handled to improve precision especially for
irregular shape spacecrafts, such as shuttles with wings.

\section{Numerical results and discussions}
\label{sect:rel and dis}

A computer program is developed to implement the technique and
test cases are run. Firstly, a collision accident between two space debris which happened On January 17, 2005 is applied.
State vectors of the two debris in the inertial system are listed in Table 1. The radius of combined geometric sphere is set to 100\,m.
The method in this paper described in Eq.~(16) and the general method described in Eq.~(5) are respectively implemented to compute
collision probability.
Figure~6 represents the variation of collision probability versus the relative position error in the along-track direction.
In the left subplot of Fig.~6, the ratio of position errors along three principle axis in the local frame is set to be 1:1:1, and 1:5:1 for the right.
We can see that the results show essential agreement between our method and the general method, with our method a little larger than the general.
Actually, the tiny deviation is caused by integral area from a circle of diameter $2r$ change to a square with the length of each side being $2r$.
But it makes no distinct difference for collision risk assessment.

\begin{center}
\textbf{\noindent\normalsize Table~1\quad State vectors in the inertial frame} \\
 [2mm]
\small
\begin{tabular}{ccccccc}
\hline
Number & $x / km$  & $y / km$  & $z / km$ & $v_x / km\cdot s^{-1}$ & $v_y / km\cdot s^{-1}$  & $v_z / km\cdot s^{-1}$ \\
\hline
07219  & -150.798  &  1182.677 & -7143.776 & 7.290 & -1.126 & -0.354 \\
26202  & -150.620  &  1182.018 & -7143.788 & 5.915 &  4.348 &  0.643 \\
\hline
\end{tabular}
\end{center}

\begin{figure}[htbp]
  \centering
  \includegraphics[width=1.0\textwidth]{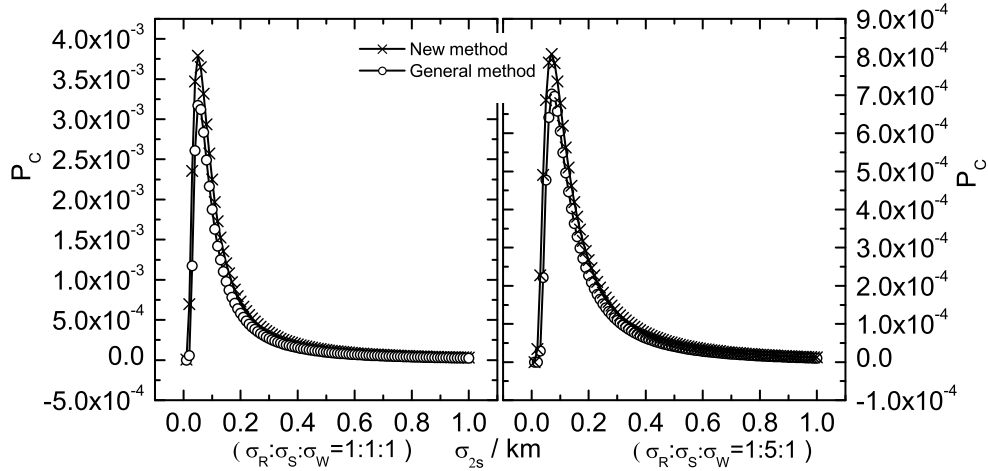}
  \caption{ \footnotesize Variation of collision probability versus the relative position error.
   The ratio of position error along three principle axis in the local frame is set to be 1:1:1 for the left subplot ,and 1:5:1 for the right.
   ( Collision probabilities about this conjunction case are also calculated in papers of others (Bai \& Chen ~\cite{bc2008}), the maximum probability corresponding to the
    right subplot is $7.12\times10^{-3}$.) }
  \end{figure}

The Center for Space Standards \& Innovation (CSSI) offers Satellite Orbital Conjunction Reports Assessing Threatening Encounters in Space (SOCRATES)
which lists the top ten of threatening encounters ordered by minimum range over the next seven days.
Based on the orbit data listed in SOCRATES on March 13, 2013 , we assess collision risk between two involved objects.
The time of closest approach and the minimum distance between the two objects are calculated, and the collision probability results derived from both our method and the general method are listed in Tab.~2. The results show that the collision probabilities from our method are on the same order of magnitude compared with the general.
Therefore, the collision probability from the method in this paper is credible to assess collision risks.

\begin{center}
\textbf{\noindent\normalsize Table~2\quad Collision probability results} \\
 [2mm]
\small
\begin{tabular}{cccccc}
\hline
Catalog  & TCA & Min & Position  & General & This\\
 Number  &( UTC ) & Range\,(km) & Error\,(km) &  Method   &  Method  \\
\hline
25415\&31445 &Mar\,18 14:44:34 & 0.115 &0.128  & 1.24E-05 & 1.52E-05 \\
20737\&20738 &Mar\,17 10:39:31 &0.104  &0.380  & 1.70E-06 & 2.15E-06 \\
27939\&31588 &Mar\,16 13:46:21 &0.098  &0.073  & 3.01E-05 & 3.51E-05 \\
11308\&32315 &Mar\,15 03:02:16 &0.094  & 0.107 & 1.80E-05 & 2.05E-05 \\
17583\&37442 &Mar\,16 14:02:50 &0.039  &0.011  & 9.39E-05 & 1.88E-04 \\
\hline
\end{tabular}
\end{center}

\section{Conclusions}
\label{sect:conclusion}
The existing general method for collision probability provides sufficiently accurate results, but does not reveal the direction connection between the collision probability and the
conjunction geometry and error covariance. Based on the geometric features of encounter, a method is developed to compute collision probability. This method gives an explicit expression to compute collision probability which is expressed as an function of the radical and the cross section components of  the relative position and error covariance. The expression is compact and advantageous to analyze the influencing factors of the collision probability directly.
 A computer program is developed to implement the technique, and the calculation results are compared with the general method. It shows that this method is applicable to most collision risk assessments in near circular orbit. Additionally, the method separates orbit motions in the radial direction from that in the cross section.
Therefore, position errors and geometric sizes of the two components can be differently handled to improve precision especially for
irregular shape spacecrafts, such as shuttles with wings. This work is under progress now.

\begin{acknowledgements}
This work was funded by the National Natural Science Foundation of China (NSFC)
under No.11203085.
\end{acknowledgements}

\label{lastpage}


\begin{thebibliography}{99}
%% you can type \apj for ApJ, \aap for A&A, \apss for Ap&SS, etc. Please consult
%% the macro chjaa.cls. You can also find them in aasguide.tex (AASTeX for ApJ, AJ, PASP)
%% Please follow the format of ChJAA's reference list

  \bibitem[2000]{aa2000} Akella M.R., Alfriend, K.T., 2000,
   Journal of Guidance, Control and Dynamics, 23, 769

  \bibitem[2008]{bc2008} Bai X.Z., Chen L., 2008, Journal of Astronautics, 29, 1435

  \bibitem[2009]{bc2009} Bai X.Z., Chen L., 2009, Chin. J. Space Sci, 29, 422

  \bibitem[1997]{ch1997} Chan K., 1997, Advances in Astronautical Sciences, 96, 1033

  \bibitem[2008]{fe2008} Francois L., Eloy S., 2008, Space Operations Communicator, 5

  \bibitem[2005]{kp2005} Kelly B.D., Picciotto S.D., 2005, AIAA paper

  \bibitem[2005]{kl2005} Klinkrad H., Alarcon J.R., Sanchez N., 2005, ESA SP-587

  \bibitem[2002]{ls2002} Leleux D., Spencer R., Zimmerman P. et al., 2002, Space OPS 2002 Conference

  \bibitem[2009]{lu2009} Liu Q.J., Chen T., Chen S.Z. et al., 2009, Advanced Measurement and Laboratory Management, 3, 21

  \bibitem[2001]{pa2001} Patera R.P., 2001, Journal of Guidance, Control and Dynamics, 24, 716

  \bibitem[2013]{sz2013} Sun R.Y., Zhao C.Y., 2013, RAA, 13, 604

\end{thebibliography}
\end{document}